\documentclass[aps,prl,twocolumn,groupedaddress,showpacs,showkeys]{revtex4-1}

\usepackage{amsmath}
\usepackage{latexsym}
\usepackage{epsfig}
\usepackage{epstopdf}
\usepackage{makeidx}
\usepackage{amsfonts}
\usepackage{mathtools}

\begin{document}

\title{Anomalous metapopulation dynamics on scale-free networks}

\author{Sergei Fedotov}
\email{sergei.fedotov@manchester.ac.uk}
\affiliation{School of Mathematics, The University of Manchester, Manchester M13 9PL, UK}
\author{Helena Stage}
\email{helena.stage@manchester.ac.uk}
\affiliation{School of Mathematics, The University of Manchester, Manchester M13 9PL, UK}

\date{\today}

\begin{abstract}
We model transport of individuals across a heterogeneous scale-free network where a few weakly connected nodes exhibit heavy-tailed residence times. Using the empirical law Axiom of Cumulative Inertia and fractional analysis we show that `anomalous cumulative inertia' overpowers highly connected nodes in attracting network individuals. This fundamentally challenges the classical result that individuals tend to accumulate in high-order nodes. The derived residence time distribution has a non-trivial U-shape which we encounter empirically across human residence and employment times.
\end{abstract}



\maketitle

\textit{Introduction.}
In the past few decades, many metapopulation models have been developed describing reaction-transport processes on scale-free networks \cite{rhodes96, grenfell98, nature-network, prl-network, Colizza08, contagion, vesp-12, human-disease, disease15, density-rate}.
The idea that the overall population can be understood as a series of spatially connected but separated `patches' \cite{classical} is useful in many areas including the migration of humans between cities \cite{Riley}, scientific collaborations \cite{barabasi02}, the spread of epidemic diseases via individual movement \cite{prl-network, holling, disease15, human-disease} and international air travel \cite{Guimera}. Often networks are assumed to be scale-free, such that the order (number of connections) of each node (patch) is drawn from a power law distribution $P(k)\sim k^{-\gamma}, \gamma>0$ \cite{network-book, barabasi99, internet-corr, barabasi09}.

While considerations of stochastic movement of individuals on a complex network are very challenging, much progress has been made using a mean-field approximation across nodes of equal order. One introduces the mean number of individuals $N_k(t)=\frac{1}{\eta_k}\sum_i\rho_{i,k}(t)$, where $\rho_{i,k}(t)$ is the number of individuals in the $i^{th}$ node of order $k$, and $\eta_k$ is the number of nodes of order $k$ \cite{prl-network, nature-network, Colizza08, vesp-12}. The equation describing transport between nodes can be written as
\begin{equation}
\frac{\partial N_k}{\partial t} = -\mathbb{I}_k(t) + k\sum_{k'}P(k'|k)\frac{\mathbb{I}_{k'}(t)}{k'},
\label{eq: start-flux}
\end{equation}
where $\mathbb{I}_k(t)$ is the flux out of a node (patch) of order $k$ and $P(k'|k)$ is the probability that a link exists from a node of order $k'$ to a node of order $k$ \cite{internet-corr, nature-network, disease15}.
Commonly it is assumed that the residence time in a node (before moving elsewhere) is exponentially distributed \cite{nature-network, prl-network, contagion}. This implies a constant escape rate $\lambda$ for which the flux is
\begin{equation}
\mathbb{I}_k(t) = \lambda N_k(t),
\label{eq: markov-flux}
\end{equation}
\cite{contagion, vesp-12, sci-rep}. The assumption of an uncorrelated network, such that $P(k'|k)=\frac{k'P(k')}{\left<k\right>}$ \cite{Newman01, internet-corr, dynamics-book}, together with Eq. \eqref{eq: markov-flux} leads to the well-known steady-state result \cite{nature-network, prl-network}
\begin{equation}
N_k^s = \frac{k}{\left<k\right>}\sum_{k'}P(k') N_{k'}^s = \frac{k}{\left<k\right>}\left<N^s\right>.
\label{eq: steady-markov}
\end{equation}
It follows from Eq. \eqref{eq: steady-markov} that the mean number of individuals in a node (patch) increases with the order. One can interpret this as individuals spending more time in well-connected nodes. This famous result has been key in developing e.g. the Page Rank algorithm and is still fundamental in our intuition regarding network behaviour.
However, such conclusions are heavily based on the assumption that the movement between patches can be approximated by a Poisson process. That is, the interval between consecutive escapes from a node (\textit{residence time}), follows an exponential probability density function (PDF) $\psi(\tau)=\lambda e^{-\lambda\tau}$. New work has emerged in recent years indicating that human activity is not Poisson distributed \cite{bettencourt06}. In particular, the efforts of Barab\'{a}si and others have demonstrated that human activity often involves heavy-tailed or Pareto type PDFs \cite{barabasi05, vazquez05,brockmann06, vazquez06, barabasi07, barabasi09,Liu10, Li12}. This is particularly relevant for human mobility due to the empirical sociological law known as `\textit{The Axiom of Cumulative Inertia}' (ACI), which suggests that the probability of a person remaining in a state increases with the associated residence time \cite{Myers67, mcginnis}. The ACI can be reformulated in terms of a power law residence time \cite{mobility} with PDF:
\begin{equation}
\psi(\tau)=\frac{\mu}{\tau+\tau_0}\left(\frac{\tau_0}{\tau+\tau_0}\right)^\mu
\label{eq: pdf}
\end{equation}
for fixed constants $\mu,\tau_0>0$.
For the \textit{anomalous} case $\mu<1$, instead of Eq. \eqref{eq: markov-flux} we obtain a fractional flux $\mathbb{I}_k^a(t)$ out of a patch
\begin{equation}
\mathbb{I}_k^a(t) = \frac{1}{\Gamma(1-\mu)\tau_0^\mu}\ _0\mathcal{D}^{1-\mu}N_k(t),
\label{eq: anom-flux}
\end{equation}
where $_0\mathcal{D}^{1-\mu}$ is the Riemann-Liouville fractional derivative defined as 
\begin{equation}
_0\mathcal{D}^{1-\mu}N_k(t)=\frac{1}{\Gamma(\mu)}\frac{d}{dt}\int_0^t\frac{N_k(\tau)}{(t-\tau)^{1-\mu}} d\tau
\end{equation}
\cite{rl-operator, klafter, angst13-1, angst16} (details in Supplementary Information and following subsection).
To the authors' knowledge no work has yet been done investigating the effect of anomalous fluxes like Eq. \eqref{eq: anom-flux} on Eq. \eqref{eq: steady-markov}, and the subsequent implications for the long-time distribution of network individuals.

So, what happens if we introduce an anomalous flux like Eq. \eqref{eq: anom-flux} into heterogenenous network models? Surprisingly, in the case of $\mu<1$, Eq. \eqref{eq: steady-markov} was radically altered beyond the effects attributable to small perturbations. Accumulation in high-order nodes did occur, but as a short-lived transient state of the network. In the long-time limit individuals aggregated in the patches with power-law residence times, invalidating Eq. \eqref{eq: steady-markov}. This \textit{fundamentally challenges the classically held belief} that individuals will tend to accumulate in the nodes of highest order \cite{nature-network, prl-network, contagion, vesp-12, disease15, human-disease}. Furthermore, these aggregated individuals exhibit a non-trivial U-shaped residence time distribution which we find to be ubiquitous across social phenomena of mobility and employment. In what follows we develop an anomalous metapopulation model describing this behaviour.\newline

\textit{Anomalous Nodes in a Network.}
We concern ourselves with transport on a heterogeneous scale-free network containing some nodes with power law distributed residence times (see Eq. \eqref{eq: pdf}), and the rest with exponentially distributed residence times. We call nodes \textit{`anomalous'} if their average residence time $\left<T\right>=\int_0^\infty\tau\psi(\tau)d\tau$ diverges. This occurs when $\mu<1$ and is the case we shall focus on (empirical evidence for its existence to follow).
We intend to show that even in the extreme case of few connections, these power law nodes are dominant in attracting network individuals. Individuals leave nodes with rates $\mathbb{T}$. For exponential residence times, $\mathbb{T}$ is constant and Eq. \eqref{eq: markov-flux} describes the flux.
Else for power law residence times, $\mathbb{T}(\tau)=\frac{\mu}{\tau+\tau_0}$ yields Eq. \eqref{eq: pdf} \cite{mobility} using $\psi(\tau)=\mathbb{T}(\tau)\exp [-\int_0^\tau\mathbb{T}(u) du]$.
The inverse residence time dependence of $\mathbb{T}(\tau)$ is another manifestation of the ACI, which we motivate as follows.
Consider a person moving to a new city: over time they develop a social circle, gain steady employment or enter family life. Consequently, the longer their residence time the more settled they become and are thus less likely to leave \cite{housingdata, motivation}.

For power law residence times it is convenient to consider the renewal measure $h(t)$. This function can be understood as the number of events per unit time, where an `event' is an individual leaving a node. $h(t)$ obeys the renewal equation $h(t)=\psi (t)+\int_{0}^{t}h(\tau)\psi (t-\tau)d\tau$ \cite{coxmiller}. One can rewrite the flux $\mathbb{I}_k(t)$ from Eq. \eqref{eq: start-flux} as
\begin{equation}
\mathbb{I}_k(t)=\frac{d}{dt}\int_{0}^{t}h(t-\tau )N_k(\tau )d\tau,
\label{eq: h-flux}
\end{equation}
which is valid for all $\psi(\tau)$ (see \cite{klafter}, Ch. 5 for the derivation). Clearly, for constant $h(t)=\lambda$ we obtain Eq. \eqref{eq: markov-flux}.
The case $\mu<1$ in Eq. \eqref{eq: pdf} requires a fractional analysis of the renewal measure, such that we obtain
\begin{equation}
h(t)=\frac{t^{-1+\mu }}{\Gamma(1-\mu )\Gamma(\mu)\tau_{0}^{\mu }}
\label{eq: renewal}
\end{equation}
as $t\to\infty$ \cite{hlimit, klafter}. Substituting Eq. \eqref{eq: renewal} into Eq. \eqref{eq: h-flux} corresponds to the anomalous fractional flux $\mathbb{I}_k^a(t)$ of Eq. \eqref{eq: anom-flux}. We will show that this flux changes the preferential residence of individuals in well-connected nodes in favour of those with anomalous flux, even if these are weakly connected. This corresponds to dominance of low-order nodes (patches) with flux $\mathbb{I}_k^a(t)$ over high-order nodes with flux $\mathbb{I}_k=\lambda N_k(t)$.
Let us for simplicity assume only anomalous nodes to have order $k_a\ll\left<k\right>$ (nodes are weakly connected). The flux $\mathbb{I}(t)$ from the balance Eq. \eqref{eq: start-flux} becomes
\begin{equation}
\mathbb{I}_k(t)=[1-\delta_{k k_{a}}]\lambda N_k(t)+\delta_{k k_{a}}\mathbb{I}_k^a(t),
\label{eq: flux-parts}
\end{equation}
where $\delta_{k k_{a}}$ is the discrete Kronecker delta. By analysis of Eq. \eqref{eq: start-flux} (details in Supplementary Information), it follows that in the limit $t\to\infty$
\begin{equation}
N_k(t)\eta_{k}\to\delta_{k k_a}N,
\label{eq: network-limit}
\end{equation}
where $N$ is the total number of individuals in the network, and $\eta_k$ the number of nodes with order $k$. \textit{Hence the anomalous nodes jointly contain all individuals as $t\to\infty$.} This key result contrasts with the popular belief that well-connected nodes are more attractive. Furthermore, similar results cannot be replicated by na\"{i}vely  introducing nodes with very low escape rates $\lambda\ll1$.

\begin{figure}
\includegraphics[scale=.7]{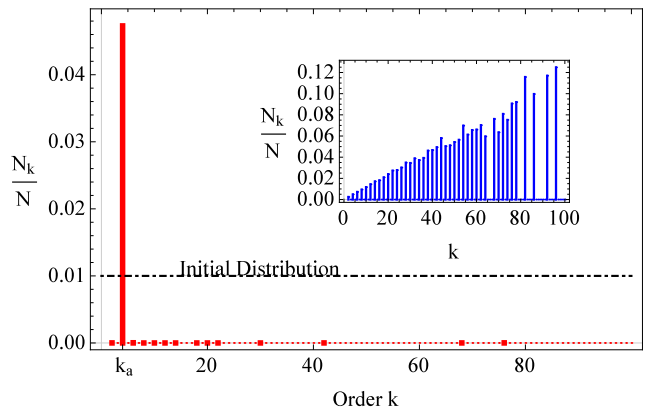}
\caption{\label{fig: power-orders}$\frac{N_k(t)}{N}$ for a network of $100$ nodes with 3 anomalous nodes, all of order $k_a=4$ with $\mu=0.5,\ \tau_0=1$, and $N=10^5$ individuals (initially distributed uniformly). Individuals eventually aggregate in the anomalous nodes. The inset shows $\frac{N_k}{N}$ if all nodes have constant escape rates $\mathbb{T}=\lambda=2$, equivalent to networks with $T(\tau)$ for $\tau_0=1,\ \mu=3$.
}
\end{figure}

We confirm the result of Eq. \eqref{eq: network-limit} by Monte Carlo simulations illustrated in Figure \ref{fig: power-orders}. A scale-free ($P(k)\sim k^{-\gamma}, \gamma=1.5,\ 2.5$), uncorrelated network was constructed using the Molloy-Reed algorithm, containing $\eta_{k_a}=3$ anomalous nodes of order $k_a=4$ \cite{mr-algorithm}. This was compared with another network where all nodes have exponential residence times and flux $\mathbb{I}_k(t)=\lambda N_k(t)$. Both simulations were carried out with $100$ nodes and $N=10^5$ individuals. Simulations were also done for networks with up to $10^4$ nodes with qualitatively similar results but a longer transient state.
\newline
Simulations almost immediately showed the individuals accumulating in nodes according to their order as described by Eq. \eqref{eq: steady-markov}. However, this behaviour was transient as the individuals then slowly moved into the anomalous nodes. We observed an initially fast rate of organisation into the classically expected configuration which then, with a (power law) slow rate, changed into a preference for the anomalous nodes. This leads to the peak in $N_k(t)$ at $k=k_a$. One can allow non-anomalous nodes of order $k_a$ in the network, though these will gradually be emptied. The only consequence is a reduced value of $N_{k_a}(t)$ as $\eta_{k_a}$ grows. Similarly, our findings are qualitatively unchanged for any $k_a>0$; this only changes how quickly accumulation occurs.

PDFs for residence times like Eq. \eqref{eq: pdf} have previously been applied to random walks \cite{sergei12, sergei13, soa15}. However, these papers do not consider the effects on a network structure, nor details pertaining to the accumulated individuals.
Related work exists considering heavy-tailed residence times in biased Watts-Strogatz networks, which demonstrated pair aggregation akin to self-chemotactic-like forcing \cite{angst13-1}. Other pattern formation on scale-free networks has been observed with order-dependent escape rates \cite{angst13}.
Patterns or dominant behaviours are known to arise in networks, either as a result of heterogeneities in $P(k)$ \cite{sci-rep} and the role of extreme values of $k$ \cite{robustness}, or following the interplay of these with escape rates or node reaction dynamics \cite{Colizza08}.\newline

\begin{figure}
\includegraphics[scale=.35]{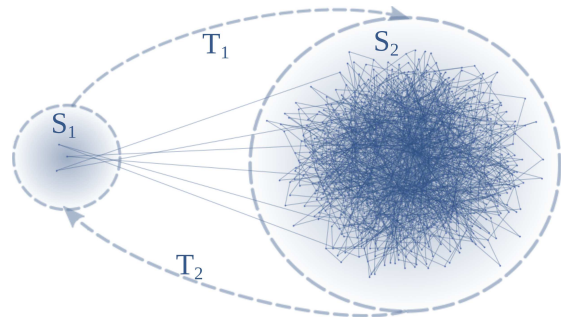}
\caption{\label{fig:states}Network separation into two states $1, 2$ with transition rates $\mathbb{T}_1,\ \mathbb{T}_2$. The exact number of nodes in each state and the number of connections between the states is insignificant, so long as $\mathbb{S}_2$ contains the majority of nodes. The intention is to demonstrate the attractiveness of $\mathbb{S}_1$, even in the extreme case where there are very few connections.
}
\end{figure}

\textit{Two-State System.}
From our simulations we observe the formation of two states in the network. There is a slow transport of individuals to the anomalous patches arising from the gradual depletion of the surrounding nodes. Consequently, we can regard this peak in individuals as one state $\mathbb{S}_1$ and the remainder of the nodes as the other state $\mathbb{S}_2$. This picture (see Figure \ref{fig:states}) allows us to find the rate at which the aforementioned peak grows.
The corresponding equations to Eq. \eqref{eq: start-flux} are
\begin{equation}
\frac{dN_{1}}{dt}=\mathbb{I}_2(t)-\mathbb{I}_1(t),\quad N_2(t)=N-N_1(t)
\label{eq: s1}
\end{equation}
where $N_i(t),\ \mathbb{I}_i(t)$ are the respective mean number of individuals in, and flux from, state $\mathbb{S}_i$. Hence the fluxes $\mathbb{S}_2\leftrightarrow\mathbb{S}_1$ in analogy to Eq. \eqref{eq: markov-flux} and Eq. \eqref{eq: h-flux} are given by $\mathbb{I}_2(t) = \lambda N_2(t)$, and $\mathbb{I}_1(t)= \frac{d}{dt}\int_{0}^{t}h(t-\tau )N_1(\tau )d\tau$ where $h(t)$ follows Eq. \eqref{eq: renewal}.
In the limit of $t\to\infty$ we neglect the derivative $dN_{1}/dt\approx0$ such that Eq. \eqref{eq: s1} becomes
\begin{equation}
N =N_1(t)+\frac{1}{\lambda}\frac{d}{dt}\int_{0}^{t}h(t-\tau)N_{1}(\tau )d\tau .
\end{equation}%
This evaluates to
\begin{equation}
N_{1}\left( t\right) =N\left( 1-\frac{h(t)}{\lambda}\right)\to N, \
N_{2}\left( t\right) =\frac{Nh(t)}{\lambda}\to 0.
\label{eq: soln-s12}
\end{equation}
as $t\to\infty$. Eq. \eqref{eq: soln-s12} thus describes the power law slow, non-stationary aggregation which is consistent with Eq. \eqref{eq: network-limit}. This phenomenon has been observed previously in other contexts \cite{shushin}, though its implications for networks has hitherto not been considered. Internal connections in $\mathbb{S}_2$ are negligible as they simply contribute slightly to the probability of remaining in $\mathbb{S}_2$ (thus increasing the time taken to aggregate in $\mathbb{S}_1$, but not the overall behaviour). 

\begin{figure}
\includegraphics[scale=.7]{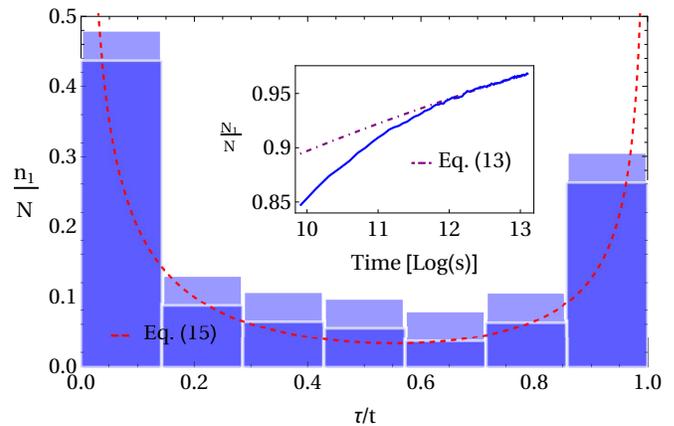}
\caption{\label{fig: data} The histogram shows $\frac{n_1(t,\tau)}{N}$, sampled from 12288 households in Milwaukee from 1950-1962. There is reasonable agreement between the data and Eq. \eqref{eq:limit-dens} for $\mathbb{T}(\tau)\approx0.55/(0.22 + \tau)$ at $t=12$ years between 1950-62. Estimated errors are indicated by the shaded regions. The inset shows $\frac{N_1(t)}{N}$ as measured from our simulations (using same parameters as Figure \ref{fig: power-orders}), thus illustrating the aggregation of individuals in $\mathbb{S}_1$ as described by Eq. \eqref{eq: soln-s12}.
}
\end{figure}

Using the same parameters, Monte Carlo simulations of the whole network were carried out to test the prediction of Eq. \eqref{eq: soln-s12} and the validity of the two-state simplification.
As shown in the inset of Figure \ref{fig: data}, the simulation is in agreement with theoretical expectations and converges to Eq. \eqref{eq: soln-s12} as $t\to\infty$. The suitability of the fit thus supports our two-state simplifying assumption.
Note that even at large times oscillations occur around the maximum, indicating that an equilibrium state does not exist.\newline

\textit{Preferential Residence.}
The aim now is to provide empirical evidence for the anomalous attractiveness of nodes with power law residence time PDFs like Eq. \eqref{eq: pdf} with $\mu<1$. Eq. \eqref{eq: soln-s12} and Figure \ref{fig: power-orders} show that individuals will tend to reside in $\mathbb{S}_1$, but what is the fine structure of these residence times? We separate the number of individuals according to their residence times. Hence $n_1(t,\tau)\Delta\tau$ gives the number of individuals with residence times in the interval $(\tau, \tau+\Delta\tau)$ with initial condition $n_1(0,\tau)=n_1^0\delta(\tau)$ where $n_1^0\ll N$. Consequently, $N_1(t)=\int_0^tn_1(t,\tau)d\tau$. We can write $n_1$ in terms of the renewal measure $h(t)$ \cite{coxmiller}
\begin{equation}
n_{1}\left( t,\tau \right) =Nh(t-\tau )\Psi (\tau ),
\end{equation}%
where the survival function $\Psi(\tau)=\int_\tau^\infty\psi(u)du=\left(\frac{\tau_0}{\tau+\tau_0}\right)^\mu$ follows from Eq. \eqref{eq: pdf}. Substituting Eq. \eqref{eq: renewal} and letting $t\to\infty$, we find a U-shaped distribution
\begin{equation}
n_{1}\left( t,\tau \right) \simeq \frac{N}{\Gamma (1-\mu )\Gamma (\mu)\tau^{\mu }\left(t-\tau\right) ^{1-\mu }}.
\label{eq:limit-dens}
\end{equation}%
This result is consistent with the generalised arc sine distributions for backward recurrence times \cite{feller} (see p.445 where $x=\tau/t$), which only holds for $\mu<1$. 

We now compare Eq. \eqref{eq:limit-dens} with empirical observations. By analysing data from an objective housing survey carried out among 16000 households in Milwaukee between 1950-1962, we obtained the residence times since moving into the current home \cite{milwaukeedata}. This was done over an interval of 12 years and allows us to `track' households and their moves as illustrated in Figure \ref{fig: data}.
The key features of the plot are the peaks in $\frac{n_1}{N}$ at $\tau\ll t$ and $\tau\sim t$, corresponding to the most likely residence times being very short or constituting the majority of the time. The same behaviour is produced by Eq. \eqref{eq:limit-dens}, and is qualitatively very different from the predictions for $\mu>1$. In the latter case where the mean residence time $\left<T\right>$ exists, one obtains the asymptotic result $n_1(t,\tau)\to\frac{\Psi(\tau)}{\left<T\right>}$ \cite{coxmiller}. This is a decaying function of residence time $\tau$ and does not provide a good description of the data in Figure \ref{fig: data}.

The presence of peaks at both low and high residence times in our data is consistent with the Axiom of Cumulative Inertia, in that most of the individuals will either be long-term residents (which do not move), or the sum of the continued in/outflux of new arrivals. We stress that these peaks only arise if $\mu<1$ is also satisfied.
Our findings are consistent with similar data obtained by the Bureau of Census during the American Housing Surveys in the period 1985-1993 \cite{Anily99}. 
Inspired by the results for human residence, the authors carried out a survey amongst permanently employed academics at The University of Manchester, and found a U-shaped distribution of employment times like Figure \ref{fig: data} (see Supplementary Information). We refer to the former case as \textit{`academic trapping'}: once a permanent position at a research institution has been found, the dynamics follow the ACI.

Our assumption that individuals follow the ACI in some nodes and not in others is used purely for the sake of simplification. To justify this, we assume now all nodes follow the ACI as given by Eq. \eqref{eq: pdf} such that some nodes are anomalous with $\mu<1$  and others have $\mu>1$. That is, all network dynamics are non-Markovian with fluxes
\begin{equation}
\mathbb{I}_k(t)=[1-\delta_{k k_{a}}]\frac{d}{dt}\int_0^t h(t-u)N_k(u)du+\delta_{k k_{a}}\mathbb{I}_k^a(t),
\end{equation}
where $\mathbb{I}_k^a$ is the fractional flux defined by Eq. \eqref{eq: anom-flux} and $h(t)$ the renewal measure for nodes with $\mu>1$.
Numerical simulations of this network qualitatively mimic Figure 1 with aggregation in the nodes with $\mu<1$.
The is understood via the mean residence time $\left<T\right>$ of the non-anomalous nodes. When $\mathbb{T}_k(\tau)=\lambda$, one finds that $\left<T\right>\approx\frac{1}{\lambda}$. Else, when $\mathbb{T}_k(\tau)=\frac{\mu_k}{\tau+\tau_0}$ and $\mu_k>1$ one obtains $\left<T\right>=\frac{\tau_0}{\mu_k-1}$. Hence, despite one treatment being Markovian and the other non-Markovian, both escape rates lead to finite amounts of time spent in the nodes. Recalling that when $\mu<1$, $\left<T\right>\to\infty$, it becomes clear why the anomalous nodes dominate the aggregation. The residence time-dependence inspired by the ACI is alone insufficient to change the qualitative behaviour of the network.\newline

\textit{Discussion and Conclusion.}
It is a commonly held belief that individuals in a scale-free network will prefer highly connected nodes (patches). 
Our work fundamentally challenges this notion when individuals' flux follows the anomalous Axiom of Cumulative Inertia as described by Eq. \eqref{eq: anom-flux}. We have shown both analytically and numerically that the flux out of anomalous nodes with power law residence times outperforms highly connected nodes in the aggregation of network individuals. We further provide empirical evidence for the associated residence time distribution $n(t,\tau)$ of aggregated individuals, motivated by the ACI \cite{motivation}.\newline
Our findings constitute an important result in the context of network theory given the wealth of evidence that human behaviour, such as our habits on web surfing and with television, follows heavy-tailed distributions \cite{barabasi07, Liu10, Li12}. Other examples of such distributions include messaging, queuing and prioritising tasks \cite{vazquez05, vazquez06, oliveira05}.

Empirical data suggests that \textit{human residence and academic employment fall into the case of anomalous behaviour}. Long durations of permanent employment lead to \textit{`academic trapping'} where dynamics obey the ACI. Despite our analysis only being valid in cases $\mu<1$, we demonstrate empirically that this is a ubiquitous example in population movement, with variations arising depending on the nature of residence (renting/owning a home). Strikingly, it is the fractional analysis of node dynamics which uncovers the essential features of our model: anomalous accumulation and a non-trivial U-shaped residence time distribution.
Our findings need not apply only to residence times in geographical regions or employed positions, but could equally describe entrenchment of ideological beliefs, convictions, etc. Owing to our model's applications to a wide range of social phenomena, we expect our findings to be of significance to a multitude of network-related human metapopulation problems.\newline

Particularly, we believe our findings will have a significant impact on network metapoulation models studying epidemiology (e.g. the SIR model) \cite{prl-network, holling, disease15, human-disease}. It is well-known that the time spent by travelers at a destination is characterised by wide fluctuations, which crucially affects the chance and duration of mixing events and therefore has a strong impact on the spread of an emerging disease \cite{tizzoni13}. We thus expect anomalous patches to be significant in understanding how diseases might spread when individuals are reluctant to leave an area. Some work on memory effects including residence time-dependence \cite{tizzoni13, ref2}, second-order Markov processes \cite{ref3} and the effects of individual movements \cite{ref1} has already been carried out. 

\begin{acknowledgments}
The authors would like to thank N. Korabel and H. Berry for useful discussions.
\end{acknowledgments}

\bibliography{biblio}
\onecolumngrid

\section*{Supplementary Information}
\subsection*{Asymptotic Results}

Let us consider Eq. (7) in more detail. If a node of order $k$ has escape rate $\mathbb{T}_k(\tau)=\frac{\mu}{\tau+\tau_0}$, we can write an intuitive description of the flux to be
\begin{equation}\tag{I}
\mathbb{I}_k(t)=\int_0^t\mathbb{T}_k(\tau)n_k(t,\tau) d\tau=\int_0^t\frac{\mu}{\tau+\tau_0}n_k(t,\tau) d\tau,
\label{eq: flux}
\end{equation}
where $n_k(t,\tau)$ is the structured density of individuals. That is, $n_k(t,\tau)\Delta t$ gives the number of individuals in a node of order $k$ with a residence times in the interval $(\tau, \tau+\Delta\tau)$. It thus follows that $N_k(t)=\int_0^tn_k(t,\tau)d\tau$ and so for a constant rate $\mathbb{T}_k$ we obtain Eq. (2). The structured density obeys the equation of motion
\begin{equation}\tag{II}
\frac{\partial n_k(t,\tau)}{\partial t}+\frac{\partial n_k(t,\tau)}{\partial \tau}=-\mathbb{T}_k(\tau)n_k(t,\tau),
\end{equation}
which we can solve using the method of characteristics to obtain  $n_k(t,\tau)=n_k(t-\tau,0)e^{-\int_0^\tau\mathbb{T}_k(u)du}$. Here, $n_k(t-\tau,0)$ is the number of new arrivals in the node of order $k$ from elsewhere. We define the exponential term to be the survival function such that $n_k(t,\tau)=n_k(t-\tau,0)\Psi_k(\tau)$. It follows from the definition that $\psi_k(\tau)=-\frac{\partial\Psi_k}{\partial\tau}=\mathbb{T}_k(\tau)\Psi_k(\tau)$. By integration we find
\begin{equation}\tag{III}
N_k(t)=\int_0^tn_k(t-\tau,0)\Psi_k(\tau)d\tau,
\label{eq: struct}
\end{equation}
which by substitution into \eqref{eq: flux} gives $\mathbb{I}_k(t)=\int_0^tn_k(t-\tau,0)\psi_k(\tau)d\tau$. By application of the Laplace transform ($\mathcal{L}_t\{f(t)\}(s)=\int_0^\infty e^{-st}f(t)dt=\widehat{f}(s)$ denotes the Laplace transformation of $f(t)$) we find
\begin{equation}\tag{IV}
\widehat{\mathbb{I}}_k(s)=\widehat{n}_{k}(s,0)\widehat{\psi}_k(s)=\frac{\widehat{\psi}_k(s)}{\widehat{\Psi}_k(s)}\widehat{N}_k(s)=s\widehat{h}_k(s)\widehat{N}_k(s),
\label{eq: lapl-flux}
\end{equation}
where we have used \eqref{eq: struct} and Eq. (4) in the last two steps. Consequently, by an inverse Laplace transformation $\mathbb{I}_k(t)=\frac{d}{dt}\int_0^t h_k(\tau)N_k(t-\tau)d\tau$ and we obtain Eq. (7).\newline

In the long-time limit of $t\to\infty$ (or equivalently $s\to0$ in Laplace space) we can find the the Laplace transform $\widehat{\psi}(s)$ of the PDF given in Eq. (4). One finds $\widehat{\psi}(s)=\left[1+(s\tau_0)^\mu\Gamma(1-\mu)\right]^{-1}$, and so the renewal measure obeys $\widehat{h}(s)=\frac{\widehat{\psi}(s)}{1-\widehat{\psi}(s)}=\left[(s\tau_0)^\mu\Gamma(1-\mu)\right]^{-1}$. By an inverse Laplace transformation we obtain Eq. (8).
Using the definition of the Riemann-Liouville operator, which for $0<\mu<1$ has the form:
\begin{equation}\tag{V}
_0\mathcal{D}^{1-\mu}N_k(t)=\frac{d}{dt}\int_0^tN_k(t-\tau)\frac{\tau^{\mu-1}}{\Gamma(\mu)}d\tau\ \to\ \mathcal{L}_t\{_0\mathcal{D}^{1-\mu}N_k(t)\}(s)=s^{1-\mu}\widehat{N}_k(s),
\end{equation}
we can (using \eqref{eq: lapl-flux} and $\widehat{h}(s)$) express the flux in terms of this quantity in the asymptotic limit. So
\begin{equation}\tag{VI}
\mathbb{I}_k^a(t)=\frac{_0\mathcal{D}^{1-\mu}N_k(t)}{\Gamma(1-\mu)\tau_0^\mu},
\end{equation}
as $t\to\infty$, which is consistent with Eq. (5).

Identifying nodes as either anomalous or not, we can substitute Eq. (9) into Eq. (1) along with the assumption of an uncorrelated network $P(k'|k)=\frac{k'P(k')}{\left<k\right>}$ to yield
\begin{equation}\tag{VII}
\frac{1}{\lambda}\frac{\partial N_k}{\partial t} = \delta_{k k_{a}}\left[\frac{k}{\left<k\right>}\sum_{k'\neq k}P(k')N_{k'}-\frac{\mathbb{I}_k^a}{\lambda}\right] +(1-\delta_{k k_{a}})\left[\frac{k}{\left<k\right>}\left(P(k_{a})\frac{\mathbb{I}_{k_{a}}^a}{\lambda}+\sum_{\mathclap{k'\neq k,k_a}}P(k')N_{k'}\right)-N_k\right].
\label{eq: network}
\end{equation}
Transforming \eqref{eq: network} into Laplace space and letting $s\to0$ (equivalent to the long-time limit $t\to\infty$) we can compare the relative values of the terms to find that $\frac{\partial N_k}{\partial t}\approx0$. Similarly, we find the dominant behaviour $\sum_{k'\neq k_a}P(k')N_{k'}(t)\gg\mathbb{I}_{k_a}^a(t)$ and $\mathbb{I}_{k_{a}}^a(t)\ll\sum_{k'\neq k,k_a}P(k')N_{k'}(t)$. Setting these terms to zero, we obtain
\begin{equation}\tag{VIII}
0 \approx \delta_{k k_{a}}\sum_{k'\neq k}P(k')N_{k'} + (1-\delta_{k k_{a}})\left[\ \sum_{\mathclap{k'\neq k,k_a}}P(k')N_{k'}-\frac{\left<k\right>}{k}N_k\right].
\end{equation}
When $k=k_a$, the mean number of individuals outside the anomalous node $\sum_{k'\neq k_a}P(k')N_{k'}=0$, and so the entire population must be present in the anomalous nodes. 
This leads to the total aggregation of individuals in nodes of order $k_a$ as described by Eq. (10).

\subsection*{Two-State Simplification}
The purpose of this section is to show that we can qualitatively approximate the long-time behaviour of the network into two states. The intention is not to prove that the overall equations exactly reduce to Eq. (11). Consider two states in \eqref{eq: network}: $k=k_a$ and $k\neq k_a$ (which we shall term $\Omega$). Hence we get
\begin{equation}\tag{IX}
\frac{\partial N_{k_a}}{\partial t} = \frac{k_a}{\left<k\right>}\lambda\sum_{k'\neq k_a}P(k')N_{k'}-\mathbb{I}_k^a(t)=\frac{k_a}{\left<k\right>}\lambda\left<N_{k'}(t)\right>_{k'\neq k_a}-\mathbb{I}_k^a(t),
\label{eq: tw-state}
\end{equation}
where $\frac{k_a}{\left<k\right>}\lambda\left<N_{k'}(t)\right>_{k'\neq k_a}$ represents the average influx from other nodes into the anomalous nodes. This approximates $\mathbb{S}_1$. Similarly, the non-anomalous nodes follow
\begin{equation}\tag{X}
\sum_{k\neq k_a}\frac{\partial N_k}{\partial t}=\frac{\partial N_\Omega}{\partial t}=\sum_{k\neq k_a}\frac{k}{\left<k\right>}\left(P(k_{a})\mathbb{I}_{k_{a}}^a(t)+\lambda\left<N_{k'}(t)\right>_{k'\neq k,k_a}\right)-\lambda N_\Omega.
\end{equation}
$\sum_{k\neq k_a}\frac{k}{\left<k\right>}P(k_a)\mathbb{I}_{k_{a}}^a(t)$ is the average anomalous flux into all the other nodes, and 
$\sum_{k\neq k_a}\frac{k}{\left<k\right>}\lambda\left<N_{k'}(t)\right>_{k'\neq k,k_a}-\lambda N_\Omega$ represents all connections in/out of order $k\neq k_a$. This approximates $\mathbb{S}_2$. As we consider most all these nodes as the state $\Omega$, these are `internal' movements in the state and thus cancel out, with the exception of any connections from nodes of order $k\to k_a$. The result is a scaling in the value of $\lambda N_\Omega$ and letting $\sum_{k\neq k_a}\frac{k}{\left<k\right>}\lambda\left<N_{k'}(t)\right>_{k'\neq k,k_a}\approx0$, which leads to the qualitatively similar Eq. (11).

\subsection*{Empirical Evidence for U-shaped Distributions}
\begin{figure}
\centering
\includegraphics[scale=1]{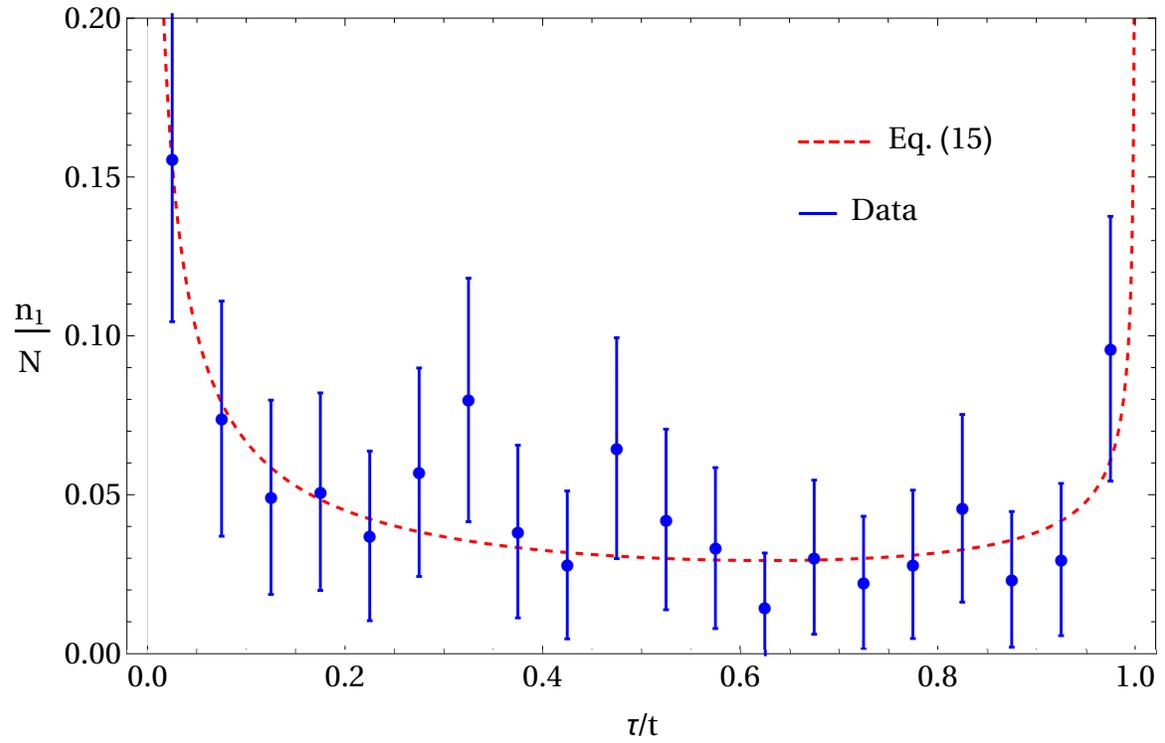}
\caption{\label{fig: maths} The distribution of employment times among 113 permanently employed academics at The University of Manchester. Variables here are $\mu=0.63,\ t=52y$. One observes the same rough qualitative behaviour as Figure 3, with peaks in the distribution at short and long employment times.}
\end{figure}

As mentioned in the paper, the appearance of U-shaped residence time distributions is expected in many areas of human mobility. To support this argument, the authors carried out a survey amongst permanently employed academic staff at The University of Manchester, the results of which are illustrated in Figure \ref{fig: maths} of this document.

The results in this figure are, due to the small sample size, not subject to rigorous statistics, and have thus not been included in the main text, but the same qualitative behaviours are observed. There are smaller peaks in the data around $\tau/t\approx 0.3,\ 0.5$, but these are attributable to the small number of available data points. Furthermore, the university underwent a merger in the year 2004, which accounts for some unusual behaviour of the data.

\end{document}